\documentclass[aps,prb,superscriptaddress,twocolumn]{revtex4-1}

\usepackage{graphicx}
\usepackage{bm}
\usepackage{amssymb}
\usepackage{float}
\usepackage{dcolumn}
\usepackage{braket}

\begin{document}

\title{Quantized current steps due to the a.c. coherent quantum phase-slip effect}

\author{R.  S.  Shaikhaidarov}
\affiliation{Royal Holloway, University of London, Egham, TW20 0EX, UK}
\affiliation{National Physical Laboratory, Hampton Road Teddington, TW11 0LW, UK}
\author{K. H. Kim}
\affiliation{Royal Holloway, University of London, Egham, TW20 0EX, UK}
\author{J. W. Dunstan}
\affiliation{Royal Holloway, University of London, Egham, TW20 0EX, UK}
\author{I. V.  Antonov}
\affiliation{Royal Holloway, University of London, Egham, TW20 0EX, UK}
\affiliation{National Physical Laboratory, Hampton Road Teddington, TW11 0LW, UK}
\author{S. Linzen}
\affiliation{Leibniz Institute of Photonic Technology, P.O. Box 100239, D-07702 Jena, Germany}
\author{M. Ziegler}
\affiliation{Leibniz Institute of Photonic Technology, P.O. Box 100239, D-07702 Jena, Germany}
\author{D. S. Golubev}
\affiliation{Pico group, QTF Centre of Excellence, Department of Applied Physics, Aalto, Finland}
\author{V. N. Antonov}
\affiliation{Skolkovo Institute of Science and Technology, Bolshoy Boulevard 30, bld. 1, Moscow, Russia 121205}
\affiliation{Royal Holloway, University of London, Egham, TW20 0EX, UK}
\author{E. Il'ichev}
\affiliation{Leibniz Institute of Photonic Technology, P.O. Box 100239, D-07702 Jena, Germany}
\author{O. V. Astafiev}
\affiliation{Skolkovo Institute of Science and Technology, Bolshoy Boulevard 30, bld. 1, Moscow, Russia 121205}
\affiliation{Royal Holloway, University of London, Egham, TW20 0EX, UK}
\affiliation{National Physical Laboratory, Hampton Road Teddington, TW11 0LW, UK}
\email{oleg.astf@gmail.com}

\begin{abstract}
The AC Josephson effect predicted in 1962 \cite{Josephson1962} and observed experimentally in 1963 \cite{Shapiro1963} as quantised {\it voltage steps} (the Shapiro steps) from photon assisted tunnelling of Cooper pairs is among the most fundamental phenomena of quantum mechanics and is vital for metrological quantum voltage standards. The physically dual effect, the AC coherent quantum phase slip (CQPS), photon assisted tunnelling of magnetic fluxes through a superconducting nanowire, is envisaged to reveal itself as quantised {\it current steps} \cite{Averin1991,Mooij06}. The basic physical significance of the AC CQPS is also complemented by practical importance in future current standards; a missing element for closing the Quantum Metrology Triangle \cite{Averin1985,Pekola2013}.
In 2012, the CQPS was demonstrated as superposition of magnetic flux quanta in superconducting nanowires \cite{astf2012}. 
However the direct sharp current steps in superconductors; the only unavailable basic effect of superconductivity to date, was unattainable due to lack of appropriate materials and challenges in circuit engineering. Here we report the direct observation of the dual Shapiro steps in a superconducting nanowire. 
The sharp steps are clear up to 26 GHz frequency with current values 8.3 nA and limited by the present setup bandwidth. The current steps have been theoretically predicted in small Josephson junctions (JJs) 30 years ago \cite{Averin1985}. However, broadening unavoidable in JJs prevents their direct experimental observation \cite{Zener,GZ}.  We solve this problem by placing a thin NbN nanowire in an inductive environment. 
\end{abstract}
\maketitle

\begin{figure}
\begin{center}
\includegraphics[width=0.45\textwidth]{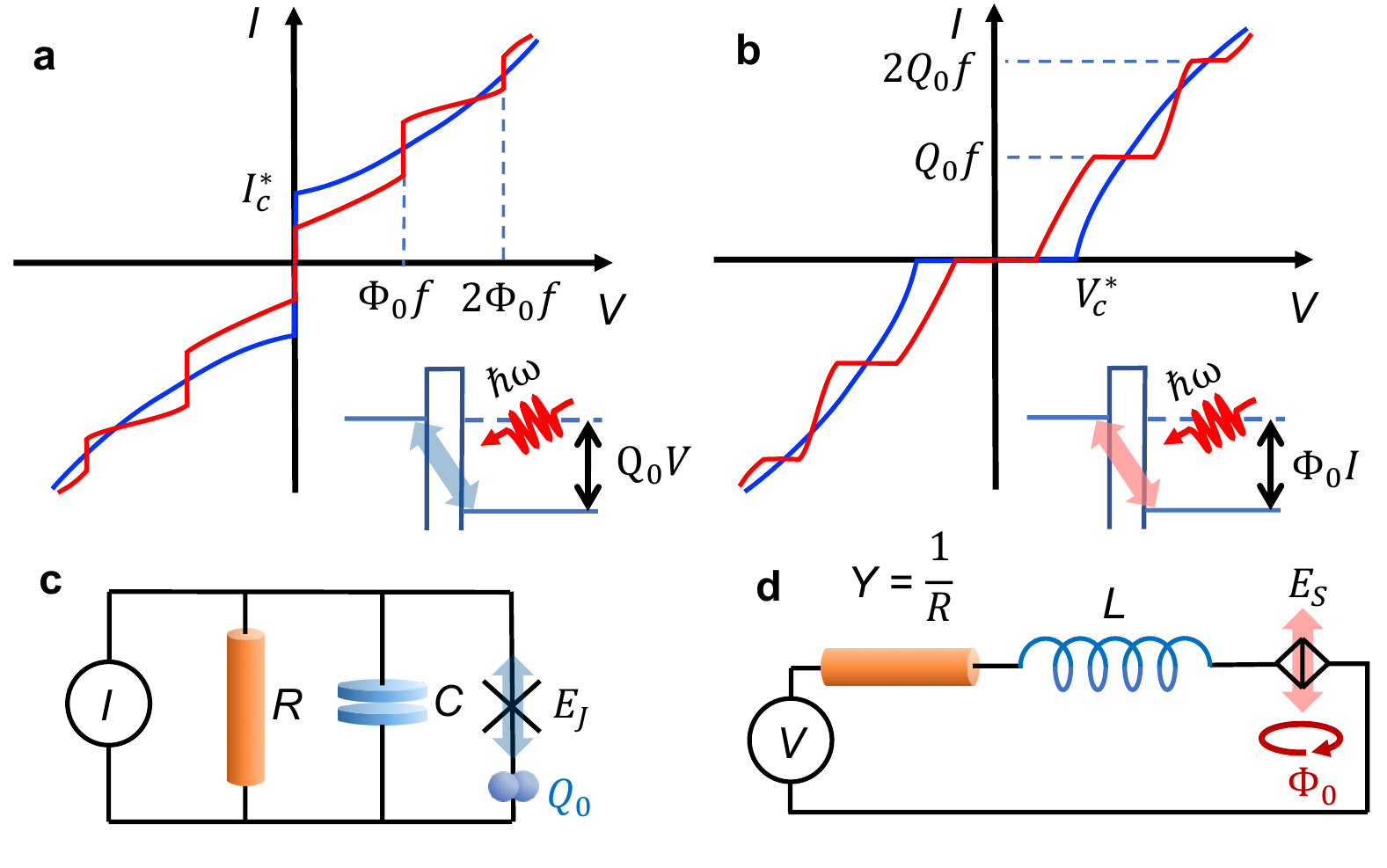}
\end{center}
\caption{\textbf{Principles of the microwave induced transport in dual circuits.} \textbf{a}, JJ transport. \textbf{b}, CQPS transport. In (\textbf{a}) and  (\textbf{b}) I-V characteristics without microwave (blue curve) and under microwave (red curve) are schematically shown. Insets show energy diagrams for the microwave assisted transport between reservoirs separated by tunnel barriers (an insulator and a nanowire) biased by $Q_0 V_{\rm dc}$ in JJ and $\Phi_0 I_{\rm dc}$ in CQPS. Effective electrical circuits for the transport measurements: \textbf{c}, for JJ; \textbf{d}, for CQPS. Tunnelling of Cooper pairs in the JJ is replaced by tunnelling of vortices through a CQPS nanowire. A capacitance $C$ and a resistor $R$ parallel to the JJ are replaced by an inductance $L$ and an admittance $Y$ in series to the CQPS junction. }
\label{fig1}
\end{figure}

Quantum-mechanical duality, a fundamental concept of physics, dictates that the phase and charge of a superconductor are quantum conjugate variables \cite{Schon1990}.
Under appropriate conditions there is an exact duality between the dynamics of both the charge, $q$, transferred through the superconductor, 
and the Josephson phase, $\varphi$, modelled as a particle motion in a $2\pi$-periodic potential.
In both cases, the non-linear effects lead to the formation of steps on the current-voltage (I-V) characteristics under microwave radiation. 
In conventional Josephson junctions (JJs), the well known Shapiro steps \cite{Tinkham1996} appear at voltages $V_n =  \Phi_0 f n$, where $f$ is the microwave frequency, $\Phi_0$ is the flux quantum and $n$ is an integer (Fig.~\ref{fig1}a). 
The dual to this is the formation of current steps in CQPS junctions. The steps form at current values $I_n=Q_0 fn$, where the charge quantum  $Q_0 = h/\Phi_0 = 2e$ 
is the charge of a Cooper pair comprised of two electrons of charge $e$ (Fig.~\ref{fig1}b) \cite{Averin1985}. 
To observe the Josephson effect, a JJ is necessarily shunted by a capacitance as shown in Fig.~\ref{fig1}c. Similarly, for the observation of CQPS, a series inductance is necessary. This also results in suppression of charge fluctuations (Fig.~\ref{fig1}d). 

Theory predicts \cite{Averin1985, Averin1991} that the dual Shapiro steps (current steps) can be observed in JJs
with resistance, $R_J$, higher than the resistance quantum, $R_Q = h/4e^{2}\approx $ 6.5~k$\Omega$,
and Josephson energy, $E_J$, close to the charging energy of the junction, $E_C$; $E_J \approx E_C$.
The Josephson energy, $E_J = I_c \Phi_0/2 \pi$, is determined by the junction critical current $I_c$,
and the charging energy, $E_C = e^2/2C$, by its capacitance $C$.
Dual Shapiro steps have been experimentally observed in the differential resistance of JJs  \cite{Kuzmin1991}.
However, the direct observation of the steps in the I-V characteristics
has proved difficult in such systems, as subsequent experiments have shown \cite{Zorin1994}.
This difficulty is likely due to hysteretic transport behaviour of junctions with
$E_J \approx E_C$  and $R_J\gg R_Q$ and the relatively small value of the energy gap separating the lowest Bloch band from the excited states. 
The gap is roughly equal to the Josephson energy $E_J = \Delta R_Q/2R_J \ll \Delta$, where $\Delta$ is the superconducting gap, facilitating Landau-Zener tunnelling to higher energy levels and promoting frequent switching between the superconducting and
the resistive states \cite{Landau, Zener}. Such processes smear the dual Shapiro steps and significantly reduce the range of bias currents and frequencies
at which they can be observed.

An alternative system where dual Shapiro steps are predicted to occur is a superconducting nanowire: a tunnelling element for magnetic fluxes \cite{GZ1,GZ,Mooij06}.
In contrast to a JJ, the nanowire may have a high value of the re-trapping current, $I_r$, 
below which the superconducting branch of the I-V characteristic is stable. 
Additionally, the energy gap in the spectrum of an ideal nanowire is very high ($\sim2\Delta$).
Moreover, CQPS in nanowires has already been used to demonstrate the superposition of flux states in different materials \cite{astf2012,Peltonen13,Peltonen16},
and for the demonstration of the interference of CQPS tunnelling amplitudes (the Aharonov-Casher effect) \cite{deGraaf2018}.

\begin{figure*}
\begin{center}
\includegraphics[width=0.9\textwidth]{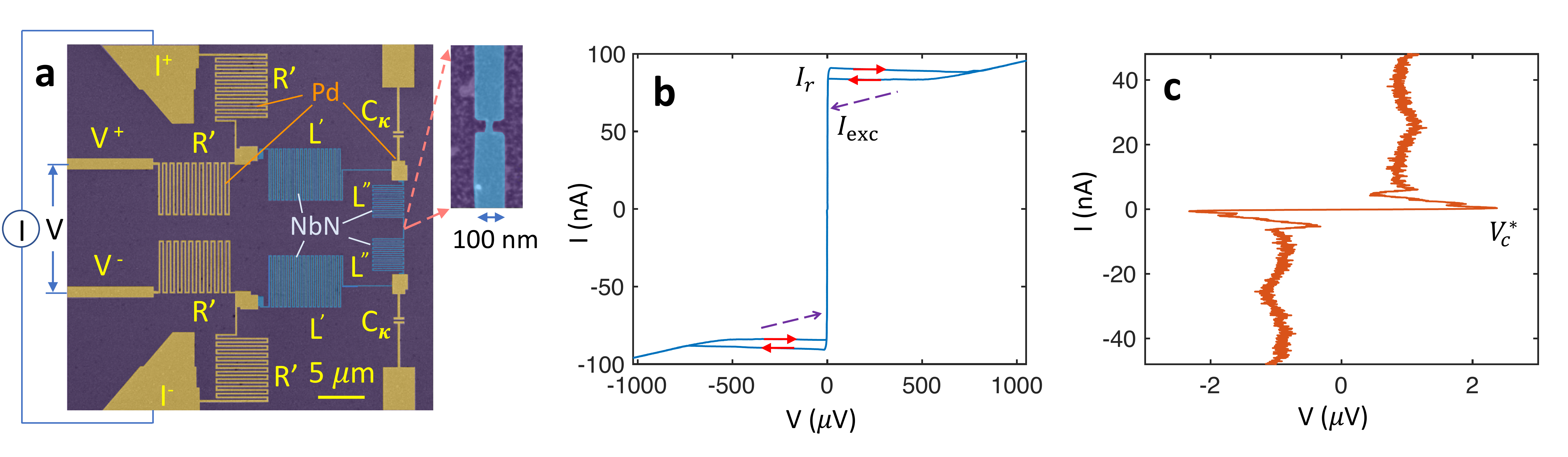}
\end{center}
\caption{\label{fig:wide} \textbf{Device and transport.} \textbf{a}, The device layout. The superconducting 100 nm wide wire with a constriction of $\sim$20$\times$50~nm$^2$ geometrical size (zoomed out) is embedded into the circuit with four compact series meandering inductances  made of the NbN films with kinetic inductances $L' \approx 1.7$~$\mu$H, $L''\approx$~0.5~$\mu$H. Inductances are connected to series Pd resistances ($R' = 11.5$ k$\Omega$) and Pd contact pads. The circuit is connected to current, $I^+/I^-$, and voltage $V^+/V^-$ leads. The microwave is delivered through an on-chip coplanar line, coupled to the circuit via capacitances $C_\kappa$. An inset shows a CQPS junction -- a small nanowire constriction. \textbf{b}, I-V characteristics in a wide voltage range demonstrate high re-trapping ($I_r$) and excess ($I_{\rm exc}$) currents. \textbf{c}, An I-V characteristic of the central part. A clear blockade is found with the re-trapping voltage $V^*_c \approx$~2.3~$\mu$V.}
\label{fig2}
\end{figure*}

Here we report the observation of distinct dual Shapiro steps in NbN nanowires under microwave drive.
The entire superconducting part of the circuit (Fig.~\ref{fig2}a) is made of a 2.7 nm thick superconducting NbN film
with critical temperature $T_c \approx 5.8$~K grown by Atomic Layer Deposition \cite{Linsen2017}, with superconducting gap in the BCS limit $\Delta \approx$~1~meV.
The film, with estimated sheet resistance $R_S \approx 4$ k$\Omega$, is close to the superconductor-insulator transition point.

The CQPS junction is a constriction with geometrical width of $\sim$20~nm and $\sim$50~nm length in a 100~nm wide NbN strip (Fig.~\ref{fig2}a).
Long 100~nm wide NbN meanders either side of the constriction
form a high kinetic inductance of total $L =2(L'+L'') \approx$~4.4~$\mu$H. 
The small geometrical size of the inductors minimises the shunting capacitance. 
Most of the devices are measured with the four-probe technique. The leads are separated from the superconducting 
structure by compact Pd resistances of total resistance $R = 2R'\approx 23.5$~k$\Omega$. 
The corresponding time constant of the system is $\tau = L/R \approx 0.2$~ns. 
A high frequency microwave signal is delivered via the two coupling capacitances $C_\kappa \approx$~1~fF made of Pd.

A typical I-V curve measured at a base temperature of 10~mK (Fig.~\ref{fig2}b) exhibits unusual features. It demonstrates distinct superconducting behaviour with an apparent critical current close to 100~nA, above which the nanowire switches to the normal state.
The large excess current $I_{\rm exc} \approx$~65~nA suggests the absence of unwanted tunnel JJs inside the nanowire 
\cite{Zaitsev1984,Abay2014}.
Conversely, blow-up in the $x$-axis reveals a current blockade at voltages $|V|<V^*_c$ with $V^*_c=2.3$~$\mu$V (Fig.~\ref{fig2}c).
We have characterised dozens of samples, some have purely superconducting behaviour with $V_c^*=0$ and others exhibiting
large critical voltages $V^*_c\ge$~50~$\mu$V \cite{deGraaf2019}. 
We have found that only samples with critical voltages in the range
$0.2$~$\mu$V $<V_c^*<$ 30 $\mu$V ($\ll \Delta/e \approx 1$~mV) exhibit the dual Shapiro steps.

\begin{figure}
\begin{center}
\includegraphics[width=0.5\textwidth]{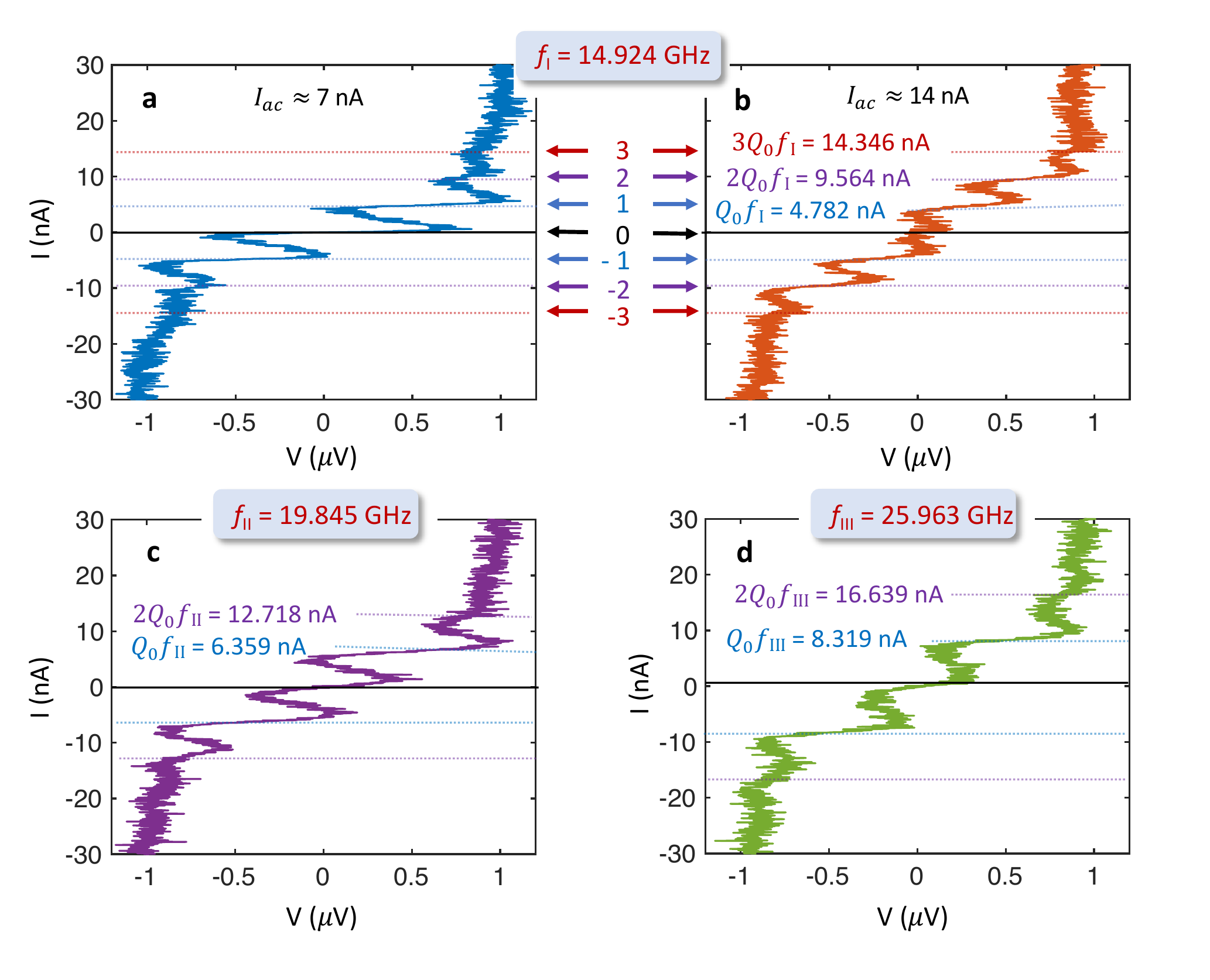}
\end{center}
\caption{\label{fig:wide} \textbf{Inverse Shapiro steps in four-probe I-V measurements.} Horizontal lines show the expected position of plateaus at $nQ_0 f$. \textbf{a}, $f_{\rm I} = $~14.924~GHz. \textbf{b}, $f_{\rm I} = $~14.924~GHz with an AC-current 2.6 times higher than in (\textbf{a}). \textbf{c}, $f_{\rm II} = $~19.845~GHz. \textbf{d}, $f_{\rm III} = $~25.963~GHz. }
\label{fig3}
\end{figure}

Under microwave excitation, current steps develop and the I-V curve is drastically modified.
Fig.~\ref{fig3} shows steps in the measured I-V characteristic at frequencies of $f_{\rm I}$~=~14.924~GHz, $f_{\rm II}$~=~19.845~GHz and $f_{\rm III}$ = 25.963 GHz. The first two steps for each frequency appear at $I_{\pm1(2)}=\pm 1(2)\times Q_0 f = \pm 2(4)\times ef$ in agreement to theory. The highest frequency at which the effect is still visible is 31~GHz ($I_1 = Q_0 f \approx 10$~nA).
The frequency is limited by the bandwidth of the transmission lines and can be further increased by optimising the setup.

In the CQPS-dominated regime, magnetic flux quanta tunnel across the nanowire with the net rate $V_{\rm dc}/\Phi_0$ proportional to the bias voltage. 
This process is dual to the charge flow of supercurrent $I^J_{\rm dc}$ ($< I^J_c$) through an insulating barrier of a JJ
with the critical current of JJs $I^J_c = 2\pi E_J/\Phi_0$. Hereafter the superscript {\it 'J'} denotes the JJ case of Figs.~\ref{fig1}a,c. In the CQPS-regime, the critical current is replaced by the critical voltage $V_c = 2\pi E_S/Q_0$, where $E_S$ is the phase-slip tunnelling energy.
The capacitance charge $Q^J = C V^J_{\rm {dc}}$ ($V^J_{\rm dc}$ is the applied dc voltage) in the circuit of Fig.~\ref{fig1}c is equivalent to 
the magnetic flux $\Phi = L I_{\rm {dc}}$ determined by DC-current $I_{\rm {dc}}$ in the circuit of Fig.~\ref{fig1}d. 
To explain the origin of this duality, we briefly summarise the theory of CQPS.
It is convenient to characterise the system state by the total number $k$ of the flux quanta $\Phi_0$, which have crossed the nanowire before a given time.
Then the adjacent states are coupled by the energy $E_S$. If a supercurrent of the form $I(t)=I_{\rm{dc}}+I_{\rm{ac}}\cos(\omega t)$, where $\omega = 2\pi f$, flows through the nanowire,
the energy of the state $\ket{k}$ becomes $E_k(t) = -k I(t)\Phi_0$.  
Hence, the system is described by a simple Hamiltonian
\begin{equation}
H = \sum_k\Big [ E_k(t) \ket{k}\bra{k} - \frac{E_S}{2} \Big ( \ket{k}\bra{ k+1} + \ket{k+1}\bra{k}  \Big ) \Big ],
\label{H}
\end{equation}
which is widely used in condensed matter physics. For example, it accurately models Bloch oscillations in semiconducting superlattices \cite{Gluek2002}.
At the resonance condition, $I_{\rm{dc}}\Phi_0 = n \hbar\omega = n Q_0 f \Phi_0$, the flux tunnelling becomes synchronised with the microwave signal and a current step is formed. This is schematically depicted in the inset of Fig.~\ref{fig1}b for $n=1$. 

The eigenstates of the Hamiltonian of Eq.~(\ref{H}) at $I(t)=0$ form the Bloch band. Each state is characterised by the quasimomentum $q$
which represents the charge transferred through the nanowire by Cooper pairs. At non-zero bias current  
this charge varies in time as $\dot q=I(t)$ and its dynamics are affected by the impedance attached to the nanowire. 
Considering the circuit of Fig.~\ref{fig1}d and assuming that the voltage bias contains 
both DC and AC components, $V(t) = V_{\rm dc} +V_{\rm ac} \cos(\omega t)$, one arrives at the following equation for $q$ \cite{Averin1985, Averin1991}
\begin{equation}
\tau\ddot \theta + \dot\theta + \omega_c\sin{\theta} = \omega_{\rm dc} + \omega_{\rm ac}\cos\omega t,
\label{LRV}
\end{equation}
where $\theta = 2\pi q/Q_0$, $\tau = L/R$, $\omega_{\rm dc} = 2\pi V_{\rm dc}/RQ_0$, $\omega_{\rm ac} = 2\pi V_{\rm ac}/R Q_0$ and $\omega_c = 2\pi V_c/R Q_0$.
According to this model, at $V_{\rm ac}=0$ and at voltages below the critical voltage, $| V_{\rm dc}| < V_c$,
the mean current $\langle \dot q \rangle = 0$. This is indeed observed in the experiment, see Fig.~\ref{fig2}c.
Outside the blockade region, Eq. (\ref{LRV}) predicts a quick drop of the time-averaged voltage $\langle V_c\sin\theta \rangle$
with increasing current \cite{Averin1985}, which is again consistent with our observations presented in Fig.~\ref{fig2}c.

Equation~(\ref{LRV}) also describes JJ dynamics of Figs.~1a,c in the so-called RCSJ model \cite{Barone1982} with the following substitutions $\theta \rightarrow \varphi$, $\tau \rightarrow C R$, $\omega_c \rightarrow 2\pi I_c R/\Phi_0$, $\omega_{\rm dc} \rightarrow 2\pi I^{J}_{\rm dc} R/\Phi_0$ and $\omega_{\rm ac} \rightarrow 2\pi I^{J}_{\rm ac} R/\Phi_0$. 
Using this analogy, one can derive a universal expression for the voltage-current characteristic of
a nanowire subject to microwave radiation,
\begin{eqnarray}
V(I)=\sum_n J_n^2\left(\frac{I_{\rm ac}}{Q_0 f} \right) V_0(I_{\rm dc} - Q_0 fn).
\label{Bessel}
\end{eqnarray}
Here $J_n(x)$ is the Bessel function
and $V_0(I)$ is the voltage-current dependence in the absence of radiation determined by the noise of the environment \cite{Averin1990}.
Eq.~(\ref{Bessel}) is the dual version of the Tien-Gordon formula \cite{TG}. The conventional Tien-Gordon formula
accurately describes the I-V characteristics of JJs
with small critical currents \cite{Rou2015, Kot2020}. Eq. (\ref{Bessel}) is universal. It is not limited to the model (\ref{LRV}) and remains
valid for any external impedance and noise provided $V_c$ is sufficiently small. 

\begin{figure}
\begin{center}
\includegraphics[width=0.47\textwidth]{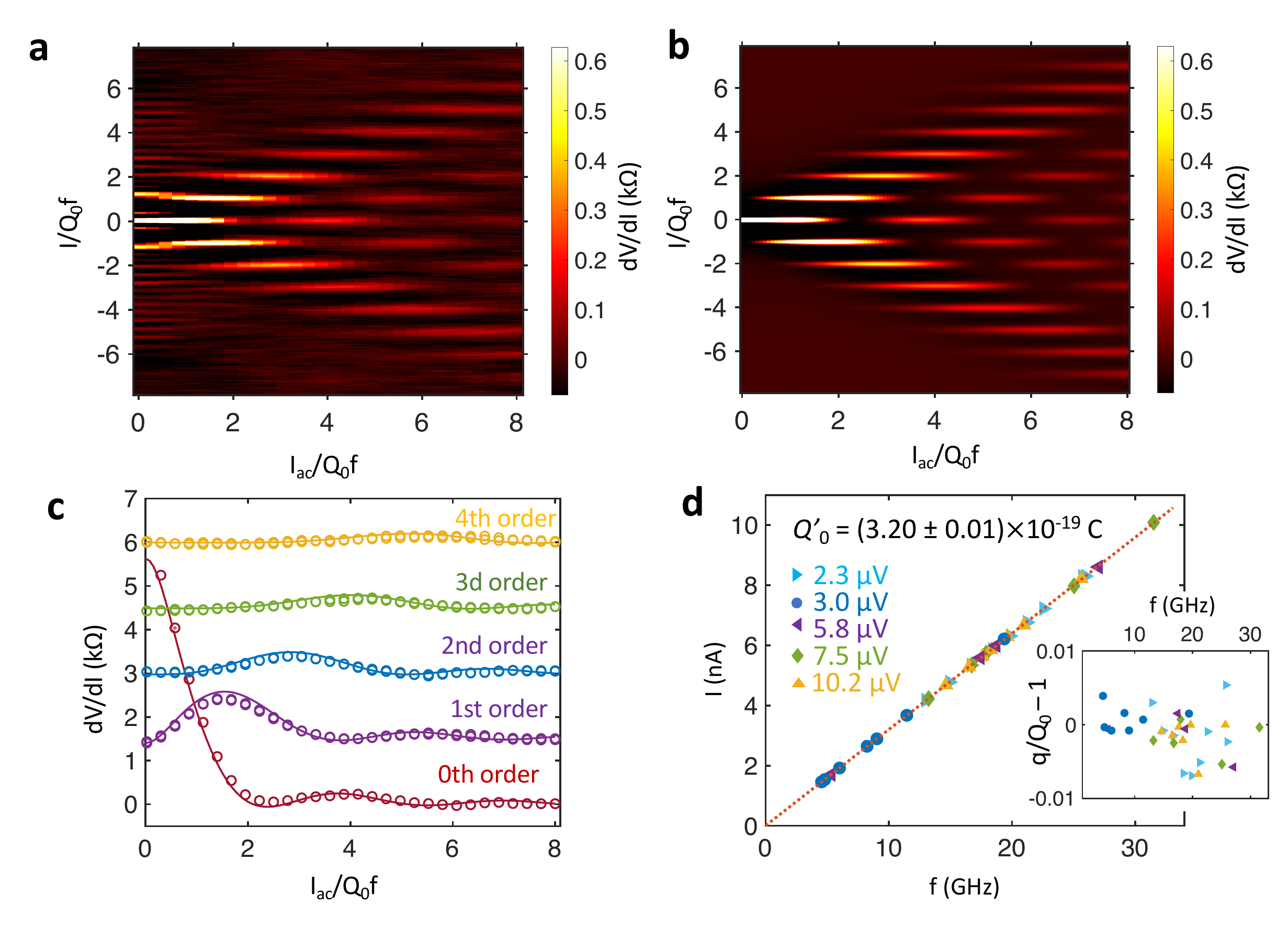}
\end{center}
\caption{\textbf{Oscillations of $dV/dI$ peaks.} \textbf{a}, $dV/dI$ characteristics in a 2D-plot experimentally measured at 14.924~GHz. \textbf{b}, Simulations, accounting heating effect from Pd resistors.
\textbf{c}, Cross-sections at positions $n$ of the quantised steps. Solid lines are simulations. Each plot is offset by $n\times 1.5~$k$\Omega$ in $y$-axis. 
\textbf{d}, $dV/dI$ for $n=\pm 1$ peak position difference calculated as $\tilde{I}_{1} = (I^{\rm max}_1-I^{\rm max}_{-1})/2$ from various samples with different $V^*_c$ (specified in the plot legend). An inset shows $q/Q_0 -1$ with $q=\tilde{I}_{1}/f$.  The red dashed line is $I =  Q_0 f$. 
}
\label{fig4}
\end{figure}

We believe that the finite slope of the current plateaus in the experiment is mainly caused
by thermal noise of the resistors, which are heated by the bias current exerting the Joule power $I^2 R \sim 10^{-12}$~W for the demonstrated steps. 
Assuming that electron-phonon coupling is the main cooling mechanism \cite{Giaz2006}, we
estimate the temperature of the resistors at the first plateaus caused by DC current in Fig.~\ref{fig3} as $T \sim $~0.2~K. 
Superconductivity is hardly expected in such a system and it is, indeed, not demonstrated as in Fig.~\ref{fig2}b. 
From the experiment we find the widths of the corresponding peaks in $dV/dI$, $\Delta I \leq$~2~nA (see Fig.~\ref{fig4}a). 
The peak width is estimated to be $\Delta I_T = \sqrt{(4 k_B T/R) \Delta f_{c}} = \sqrt{k_B T/L}$. In our $LR$ circuit $\Delta f_{c} = R/4L$, which gives $\Delta I_T \approx$~1~nA. In the discussed sample, the experimentally observed peaks at non-zero bias are typically close but slightly wider than $\Delta I_T$. The widths mainly depend on the AC-driving current amplitude.

Simulations based on Eq. (\ref{Bessel}) are presented in Fig.~\ref{fig4}b. We have
fitted the dependence of the differential resistance on the bias current without microwave power
to the form $dV/dI=A(\Delta I^2 -I_{\rm dc}^2)/(\Delta I^2 + I_{\rm dc}^2)^2$ expected for the blockade region \cite{Averin1990},
and computed $dV/dI$ at non-zero powers from Eq. (\ref{Bessel}) adjusting the peak width according to $\Delta I = \Delta I_0 (1  +\alpha I_{\rm ac}^2 )^{0.1}$, where $\Delta I_0$ and $\alpha$ are taken to reproduce experimental $\Delta I$ at zero and maximal $I_{\rm ac}$. The dependence comes from the assumption of phonon cooling, when $\Delta I \sim \sqrt{T_{eff}}$ due to the effective temperature $T_{eff} = (T_0^5 + W_{\rm ac})^{1/5}$  \cite{Giaz2006}, where the AC power is $W_{\rm ac}\sim I_{\rm ac}^2$. 
Figure~\ref{fig4}c shows the oscillations of $dV/dI$ with the amplitude of the applied microwave signal
at the current steps, i.e. at $I_{\rm dc} = n Q_0 f$ (dots), together with the simulations shown by solid lines. 
These oscillations are primarily determined by the Bessel functions of Eq.~(\ref{Bessel}) with an additional decay caused by temperature rise at increasing power.

In Fig.~\ref{fig4}d we plot the positions of the peaks in $dV/dI$ versus frequency for several samples with different critical voltages $V_c$. 
We fit these data with $I_{\rm dc}=Q'_0f$ and obtain $Q'_0 = (3.20\pm0.01)\times 10^{-19}$~C, which agrees with the Cooper pair charge $2e$. 
In the inset we plot the ratios $I_{\rm dc}/Q_0 f$ for all data points to reveal the scattering of the data points. 

In addition to the fundamental nature of the phenomenon, the observation of the dual Shapiro steps will have an impact on metrology. Similarly to the ordinary Shapiro steps used for the commercial voltage standards, the dual Shapiro steps can be utilised for the quantum current standards.
In 2019 the Metrological World Congress announced the new standard of the electric current $I$ defining it as a product of frequency and the charge quantum.
So far, quantum standards exist for voltage $V$ and resistor $R$ and to close the 'Metrology Triangle' (in one of its definitions) with three interrelated quantities $V$-$R$-$I$x, the direct quantum current standard is required.
Currently, there exist several devices exploiting transfer of individual electrons (not on coherent quantum phenomena),
which realise such a primary standard \cite{Pekola2013}. The most advanced one is the single electron pump,
where the electrons are transferred through a semiconducting quantum dot \cite{Giblin2020}.
To date, a maximum current of 170 pA has been reached.
Further increase of the current is limited by non-adiabatic excitation of electrons to the higher levels in the dot, which also limits the accuracy of the device.
We demonstrate a larger value of the current at the first plateau approaching 10~nA and it can be further increased. Importantly, there is also room for optimizing noise and filtering, which will result in flatter current plateaus. 

\textbf{Acknowledgments}

This work was supported by European Union’s Horizon 2020 Research and Innovation Programme under Grant Agreement No. 862660/QUANTUM E-LEAPS, Engineering and Physical Sciences Research Council (EPSRC) Grant No. EP/T004088/1.

\textbf{Author Contributions}

O.V.A. proposed, simulated and planed the experiment and circuit design, analysed data. R.S.S. designed and fabricated various samples, planned the experiment, analysed data. R.S.S. made the experiments with a significant contribution of K.H.K., J.W.D., I.V.A. V.N.A. K.H.K. designed and fabricated samples, analysed data, prepared figures. 
S.L., M.Z., E.V.I. developed technology and M.Z., S.L. fabricated NbN films. D.S.G. provided theory and simulations of the experiment, wrote the manuscript. O.V.A., E.V.I., D.S.G., V.N.A. wrote the manuscript. 

\bibliography{AC_CQPS_arxiv.bib}

\end{document}